\documentclass[twocolumn]{aastex62}
\usepackage{amsmath}
\graphicspath{{./}{figures/}}

\shorttitle{Identifying Trapped Interstellar Objects}

\begin{document}

\title{Identifying Interstellar Objects Trapped in the Solar System through Their Orbital Parameters}

\email{amir.siraj@cfa.harvard.edu, aloeb@cfa.harvard.edu}

\author{Amir Siraj}
\affil{Department of Astronomy, Harvard University, 60 Garden Street, Cambridge, MA 02138, USA}

\author{Abraham Loeb}
\affiliation{Department of Astronomy, Harvard University, 60 Garden Street, Cambridge, MA 02138, USA}




\begin{abstract}

The first interstellar object, `Oumuamua, was discovered in the Solar System by Pan-STARRS in 2017, allowing for a calibration of the abundance of interstellar objects of its size and an estimation of the subset of objects trapped by the Jupiter-Sun system. Photographing or visiting these trapped objects would allow for learning about the conditions in other planetary systems, saving the need to send interstellar probes. Here, we explore the orbital properties of captured interstellar objects in the Solar System using dynamical simulations of the Jupiter-Sun system and initial conditions drawn from the distribution of relative velocities of stars in the Solar neighborhood. We compare the resulting distributions of orbital elements to those of the most similar population of known asteroids, namely Centaurs, to search for a parameter space in which interstellar objects should dominate and therefore be identifiable solely by their orbits. We find that there should be thousands of `Oumuamua-size interstellar objects identifiable by Centaur-like orbits at high inclinations, assuming a number density of `Oumuamua-size interstellar objects of $\sim 10^{15} \; \mathrm{pc^{-3}}$. We note eight known objects that may be of interstellar origin. Finally, we estimate that LSST will be able to detect several hundreds of these objects.

\end{abstract}



\section{Introduction}

The first interstellar object, `Oumuamua, was discovered in the Solar System by the Pan-STARRS telescope \citep{Meech2017, Micheli2018}. Several follow-up studies of `Oumuamua were conducted to better understand its origin and composition \citep{Bannister2017, Gaidos2017, Jewitt2017, Mamajek2017, Ye2017, Bolin2017, Fitzsimmons2018}. The detection of `Oumuamua allowed for a calibration of the number density of objects of similar size, estimated to be $\sim 10^{15} \; \mathrm{pc^{-3}}$ \citep{Do2018}. This updated number density is much higher than the previous estimate of $\lesssim 10^{12} \; \mathrm{pc^{-3}}$ \citep{Moro-Martin2009, Engelhardt2017}. \citet{Lingam2018} used this calibration to estimate the capture rate of `Oumuamua-size interstellar objects by means of gravitational interactions with Jupiter and the Sun to be $1.2 \times 10^{-2} \; \mathrm{yr^{-1}}$, and the resulting number of `Oumuamua-size interstellar objects bound to the Solar System at any given time to be $\sim 6 \times 10^{3}$. Observing or visiting such objects could allow searching for signs of extraterrestrial life locally, without the need to send interstellar probes \citep{Loeb2018}. However, such a search would come with a caveat that most asteroids in the Solar System reside outside the habitable zone. Here, we explore whether it is possible to identify trapped interstellar objects through their orbital parameters alone.

The outline of the paper is as follows. In Section~\ref{sec:methods} we explore the orbital parameters of trapped interstellar objects by simulating their gravitational capture by the Jupiter-Sun system. In Section~\ref{sec:results} we describe our results, and in Section~\ref{sec:trapped} we explore their implications in identifying interstellar objects in our solar system through their orbital parameters. In Section~\ref{sec:candidates} we identify known objects that may be of interstellar origin, and in Section~\ref{sec:detect} we investigate the future detectability of the trapped interstellar object population. Finally, Section~\ref{sec:conclusion} summarizes our main conclusions.

\section{Simulation Methods}
\label{sec:methods}
In this simulation, we consider interactions only with Jupiter, since it is the most massive planet. To model the motion of interstellar objects under the gravitational influence of the Jupiter-Sun system, we developed Python code that randomly initializes and integrates the motions of particles from their points of closest approach to Jupiter to both the past and the future, and searches for particles that are initially unbound yet end in bound orbits after their gravitational interaction with Jupiter. The Python code created for this work used the open-source N-body integator software \texttt{REBOUND}\footnote{https://rebound.readthedocs.io/en/latest/} to trace the motions of particles under the gravitational influence of the Jupiter-Sun system \citep{Rein2012}.

We initialize the simulation with the Sun, Jupiter, and a volume of test particles surrounding Jupiter at their distances of closest approach to the planet. The Sun and Jupiter define the ecliptic plane. We set the range of impact parameter, $b$, of the particles relative to Jupiter to be between 1 and $10 \; \mathrm{R_{J}}$, where $R_{J}$ is Jupiter's radius. The upper limit of $10 \; \mathrm{R_{J}}$ was chosen after runs of the complete code with $b$ set to vary between 0 and $\sim 10^3 \; \mathrm{R_{J}}$ resulted in no captured particles for $b > 10 \; \mathrm{R_{J}}$. To choose the value of $b$ for each particle, we draw randomly from a weighted distribution in which the likelihood of a particle having any given value of $b$ between 1 and $10 \; \mathrm{R_{J}}$ is proportional to $b^2$, due to the fact that each value of $b$ represents a spherical infinitesimal shell of volume $4\pi b^2 \; db$ around Jupiter.
	
For each value of $b$ that is chosen, we randomly pick an angle within the ecliptic plane between 0 and $2\pi$, as well as a zenith angle between 0 and $\pi$. Using these two angles, we set the direction of each particle's velocity vector.

We randomly draw the initial speed of each particle, $v_{\infty}$, from a Maxwellian distribution with velocity dispersion 40 km/s, since this is the approximate characteristic distribution of the velocity dispersion for stars in the Solar neighborhood \citep{Binney2008,Li2016}. From the constructed velocity vector, we subtract the motion of the Sun relative to the LSR, $\mathrm{(U, V, W)_{LSR}= (11.1, 12.2, 7.3)\;}$ $\mathrm{km \: s^{-1}}$ \citep{Schonrich2010}. Finally, we pick a random position in Jupiter's orbit and account for its velocity vector. We subsequently calculate the speed of each particle using conservation of energy at closest approach, using the equation,

\begin{equation}
\mathbf{E_{k} = E_{k,\infty} - E_{g,\odot} - E_{g, Jupiter} \; \; .}
\end{equation}
While the gravitational influence of the Sun is not very large at the distance of Jupiter, we have included it in our calculation of each interstellar object's velocity vector. The resulting velocities approximate the characteristic velocity of interstellar objects, however any other object with the same velocity, such as an Oort cloud object ejected at a particular speed, could be misconstrued with interstellar objects in this analysis. 

To ensure that each particle is at its distance of closest approach, we require the position vector to lie in the plane perpendicular to the velocity vector relative to Jupiter. For each particle, we pick a random angle between 0 and $2\pi$ to determine in which direction the position vector points within this plane. Using the impact parameter, $b$, and the angle within the plane perpendicular to the velocity vector, we construct each particle's position vector. At this point, we have fully initialized each particle with both a position and a velocity vector.

In the first stage of the simulation, we integrate all of the particles backward in time and determine which ones are initially unbound. We use the \texttt{WHFast}\footnote{https://rebound.readthedocs.io/en/latest/ipython/WHFast.html} integrator in \texttt{REBOUND} to trace each particle from $t = 0$ to an earlier time $-t_i$ \citep{Rein2015}, where $t_i$ is an amount of time to sample either side of the closest approach to Jupiter. The only constraint on $t_i$ is that it is a time interval at and above which the results of the simulation do not change, on the order of Jupiter's orbital period. We then compute the escape velocity relative to the Jupiter-Sun system for each particle using the expression below and label all particles with speed $v(-t_i) > v_{esc}$ as initially unbound, where

\begin{equation}
v_{esc} = \sqrt{ \frac{2GM_{\odot}}{d_{\odot}} + \frac{2GM_{J}}{d_{J}}} \; \; ,
\end{equation}
with $d_{\odot}$ and $d_{J}$ being the distances from the Sun and Jupiter, respectively. In the second stage of the simulation, we integrate the particles with unbound initial conditions forward in time and find which ones will end in bound orbits around the Sun. We use \texttt{WHFast} to integrate each particle from $t = 0$ to $t_i$. We again compute each particle's escape velocity and determine that all particles with speed $v(t_i) < v_{esc}$ become gravitationally bound and therefore satisfy the condition of capture.

Finally, our Python code calculates initial conditions and final orbital parameters for the captured particles. At $-t_i$, we calculate each particle's incoming zenith angle, $\theta$, where $90^\circ$ corresponds to the plane of the ecliptic. We then use Orbital\footnote{https://pypi.org/project/OrbitalPy/}, an open source orbital mechanics package, to compute each particle's semi-major axis $a$,  eccentricity $e$, and inclination $i$, at $t_i$. We subsequently compute orbital period, $T$, perihelion distance, $q$, and aphelion distance, $Q$. 

\section{Simulation Results}
\label{sec:results}

\begin{figure}
  \centering
  \includegraphics[width=.9\linewidth]{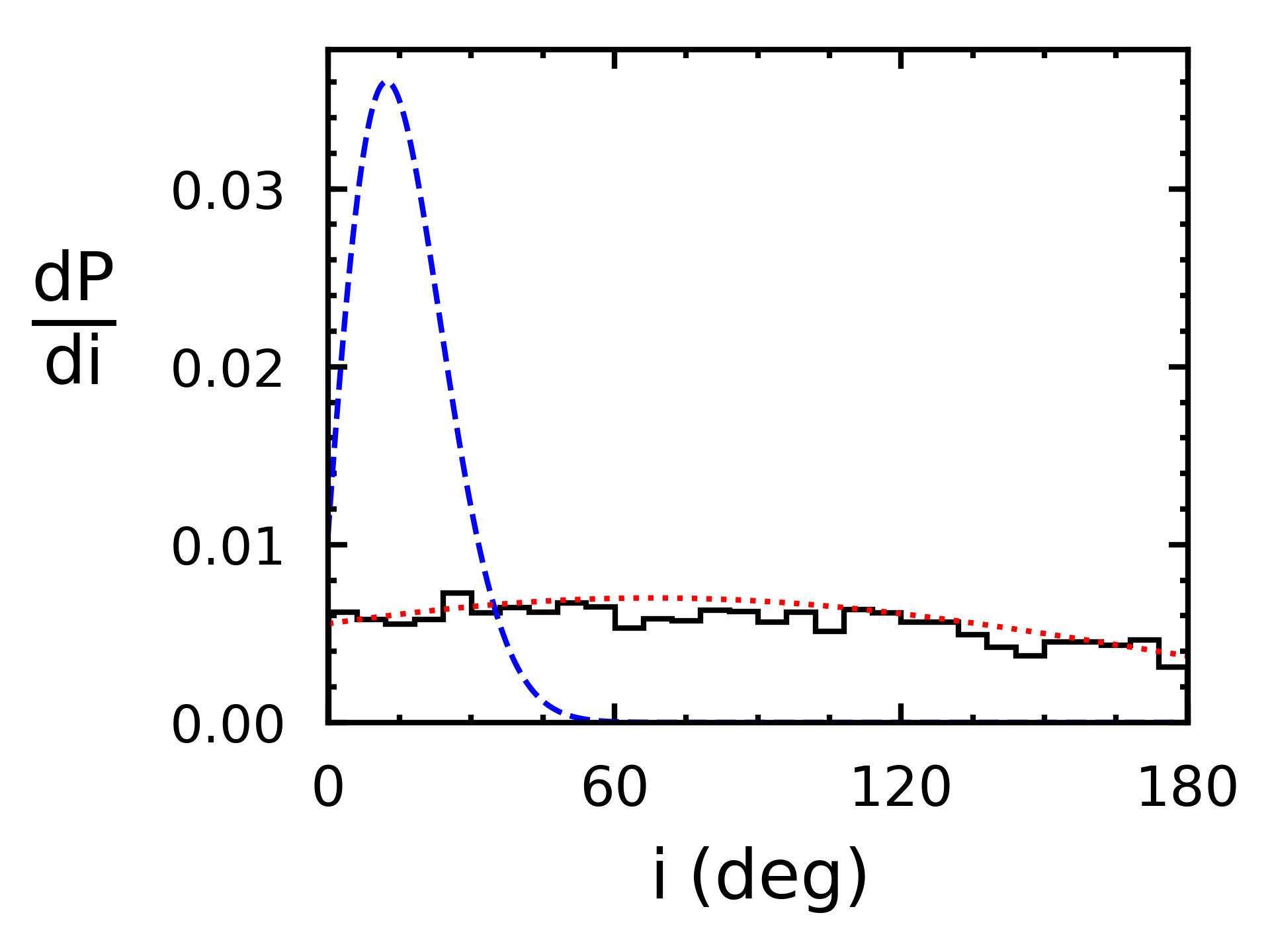}
    \caption{Distribution of inclination, $i$, of captured objects. The blue dashed line shows $P_{Centaur}(i)$, using equation (\ref{eq:4}), and the red dotted line shows $P_{interstellar}(i)$, using equation (\ref{eq:7}).}
    \label{fig:fig2}
\end{figure}

We ran our Python code for $10^5$ particles, which resulted in 3202 instances of capture. We verified that the percentage of interstellar objects captured by Jupiter remained at $\sim$3\% for $2\times$ and $5\times$ particles. The distribution of semi-major axis, $a$, is positively skewed with a median of 13 AU, $25^{\text{th}}$ and $75^{\text{th}}$ percentiles of 4.3 AU and 16 AU, respectively, and a range of 0.93 AU to $\sim 5 \times 10^{3}$ AU. The distribution of eccentricity, $e$, is negatively skewed with a median of 0.74, $25^{\text{th}}$ and $75^{\text{th}}$ percentiles of 0.53 and 0.89, respectively, and a range of up to $\sim 1$. The distribution of inclination, $i$, is positively skewed with a median of $82^\circ$, $25^{\text{th}}$ and $75^{\text{th}}$ percentiles of $41^\circ$ and $124^\circ$, respectively, and a range of up to $180^\circ$. The probability distribution of inclination, $i$, is displayed in Fig.~\ref{fig:fig2}, and is normalized to unit area. Owing to the nearly flat distribution of inclinations in Fig.~\ref{fig:fig2} that results from the broad initial velocity distribution, the population of interstellar objects dominates over the Centaur inclination distribution at an inclination angle of $48^\circ$.

\section{Identifiability of Trapped Interstellar Objects}
\label{sec:trapped}
Is it possible to identify trapped interstellar objects through their orbital parameters alone? To explore this question, we compare our orbital element distributions of trapped $\sim \mathrm{100 \; m}$ size interstellar objects to those of $\sim \mathrm{100 \; m}$ size background objects which originated in the Solar System. The population of Solar System objects with orbital elements similar to our theoretical population are the Centaurs, which are defined by \citet{Grav2011} as objects that have $5.5 \leq a \leq \mathrm{30\;AU}$ and $q > \mathrm{5.5\;AU}$. We will call our theoretical population `Centaur-like' as the objects have $\sim 3 \leq a \leq \mathrm{30\;AU}$ and $q > \mathrm{2\;AU}$. 

The size distribution of Centaurs is not well-constrained, and $\sim \mathrm{100 \; m}$ diameter Centaurs have yet to be discovered. We will estimate the number density of $\sim \mathrm{100 \; m}$ size Centaurs by using a model derived by \citet{Bauer2013}: $\: n(>D) \propto D^{-1.7}$, where $D$ is the diameter of the object, since size distribution tends to follow a power law distribution for a given population \citep{Trilling2017}. Extrapolating this model to $D \sim \mathrm{100 \; m}$ gives us an estimate of $\sim 10^{7}$ Centaurs.

\citet{Jedicke1997} describe the distributions of $a$, $e$, and $i$ for Centaurs using analytic expressions. We normalize these expressions, as corrected by \citet{Grav2011}, to the respective ranges for each parameter, forming the following probability distributions:
\begin{equation}
P_{Centaur}(a) = \frac{1}{6.9\sqrt{2\pi}} \: \exp \: \lbrack-\frac{\left(a- \mathrm{32\;AU}\right)^2}{2\cdot\left(\mathrm{6.9\;AU}\right)^2}\rbrack \; \mathrm{AU^{-1}} \; \; ,
\label{eq:2}
\end{equation}
\begin{equation}
P_{Centaur}(e) = 2.26 \: \exp \: \lbrack-\frac{\left(e-0.21\right)^2}{2\cdot\left(0.21\right)^2}\rbrack \; \; ,
\label{eq:3}
\end{equation}
\begin{equation}
P_{Centaur}(i) = 3.96 \times 10^{-3} \: (i+2.7) \: \exp \: \lbrack-\frac{\left(i+2.7^\circ \right)^2}{2\cdot\left(15^\circ \right)^2}\rbrack \; \mathrm{deg^{-1}} \; \; ,
\label{eq:4}
\end{equation}

To mathematically describe the orbital element distributions of trapped interstellar objects, we fit Gaussian distributions to the segments of distributions that deviate most significantly from their Centaur counterparts. We describe $0 \leq a \leq \mathrm{10\;AU}$, $0 \leq e \leq 1$, and $0 \leq i \leq 180^\circ$ with Gaussians, and express the resulting functions as normalized probability distributions:
\begin{equation}
P_{interstellar}(a) = 1.5 \times 10^{-1}  \: \exp \: \lbrack-\frac{\left(a-\mathrm{4\;AU}\right)^2}{2\cdot\left(\mathrm{3\;AU}\right)^2}\rbrack \; \mathrm{AU^{-1}}  \; \; ,
\end{equation}
\begin{equation}
P_{interstellar}(e) = 2.2 \: \exp \: \lbrack-\frac{\left(e-1\right)^2}{2\cdot\left(0.36\right)^2}\rbrack \; \; ,
\end{equation}
\begin{equation}
P_{interstellar}(i) = 6.4 \times 10^{-3} \: \exp \: \lbrack-\frac{\left(i-68^\circ \right)^2}{2\cdot\left(100^\circ \right)^2}\rbrack \; \mathrm{deg^{-1}} \; \; .
\label{eq:7}
\end{equation}

The total number of $\sim \mathrm{100 \; m}$ size interstellar objects is estimated to be $\sim 6 \times 10^3$  \citep{Lingam2018}, whereas the total number of $\sim \mathrm{100 \; m}$ size Centaurs is estimated to be $\sim 10^{7}$.  The best way to distinguish between the populations is through their orbital inclination. We derive $N_{Centaur}(i)$ and $N_{interstellar}(i)$ as follows:

\begin{align}
&N_{Centaur}(i) = 10^7 \cdot P_{Centaur}(i) \; \; , \\
&
N_{interstellar}(i) = 6 \times 10^3 \cdot P_{interstellar}(i) \; \; .
\end{align}

For the number of interstellar objects to dominate over Centaurs, we require that $\frac{N_{interstellar}(i)}{N_{Centaur}(i)} > 10$. We find that this condition holds for $i \gtrsim 48^\circ$, and we integrate $N_{interstellar}(i) \; di$ from $48^\circ$ to $180^\circ$, resulting in an expected identifiable trapped population of $N_{interstellar} \sim 4000$.

\section{Trapped Interstellar Object Candidates}
\label{sec:candidates}
\begin{table}
	\centering
	\caption{Orbital parameters for possible trapped interstellar objects, namely semi-major axis, $a$, eccentricity, $e$, inclination, $i$, orbital period, $T$, perihelion distance, $q$, and aphelion distance, $Q$.}
	\label{tab:1}
	\resizebox{\linewidth}{!}{\begin{tabular}{lccccccc}
        \hline
        \textbf{Candidate} & \textbf{a (AU)} & \textbf{e} & \textbf{i (deg)} & \textbf{P (yr)} & \textbf{q (AU)} & \textbf{Q (AU)} & \textbf{H}\\
        \hline
        2018 WB1           & 8.12            & .73        & 152.1          & 23.13            & 2.23            & 14.01            & 17.0            \\
        2013 YG48           & 8.19            & .75        & 61.3          & 23.44            & 2.03            & 14.35            & 17.2            \\
        2018 TL6           & 1.72            & .79        & 170.9          & 23.76            & 1.72            & 14.80            & 19.9            \\
        2017 SV13           & 9.65            & .79        & 113.2          & 29.99            & 2.01            & 17.30            & 18.2            \\
        2013 JD4           & 12.25            & .87        & 73.0          & 42.85            & 1.63            & 22.86            & 16.5            \\
        2018 AS18           & 13.01            & .87        & 63.0          & 46.94            & 1.66            & 24.37            & 16.0            \\
        2008 WA95          & 14.05           & .88        & 60.1          & 52.68            & 1.72            & 26.38            & 16.8            \\
        2016 WS1           & 14.43           & .88        & 53.0           & 54.80           & 1.70            & 27.16            & 17.3  \tabularnewline         
        \hline
    \end{tabular}}
\end{table}

While the size distribution for interstellar objects is yet unconstrained, we will use the size distribution of Centaurs as an estimate, namely ${n(>D) \propto D^{-1.7}}$, where $D$ is the diameter of the object, as Centaurs are the most dynamically similar population. Approximating ${n(>D) \sim 6 \times 10^3}$ for ${D = \mathrm{100 \; m}}$ \citep{Lingam2018}, we find that ${n(>D) \gtrsim 10}$ for ${D \gtrsim \mathrm{4 \; km}}$.

We assume a very low albedo (.05) to find a conservative cutoff value for absolute magnitude of 4 km size objects, $H > 16$. We then filter all objects in the MPC database to those with values of semi-major axis, $a$, and eccentricity, $e$, within our established $25^{\text{th}}$ and $75^{\text{th}}$ percentile bounds, as well as inclination, ${i \gtrsim 48^\circ}$, and absolute magnitude, $H > 16$. We set 5.5 AU as the lower bound for semi-major axis, $a$, since our analysis only includes Centaurs. We find eight potential trapped interstellar objects, enumerated in Table~\ref{tab:1}.

\section{Future Detectability of Trapped Interstellar Objects}
\label{sec:detect}

Future surveys, such as the Large Synoptic Survey Telescope (LSST\footnote{https://www.lsst.org/}), will be able to search for our predicted population of trapped interstellar objects. While the Hyper Suprime-Cam Survey (HSCS\footnote{https://hsc.mtk.nao.ac.jp/ssp/survey/}) has comparable sensitivity, with the capability of reaching a magnitude of $24.5$ with a 10-to-1 signal-to-noise ratio, LSST has a far larger field of view, covering 38 times more area on the sky, which will be more advantageous for the search. While HSC could reach a magnitude of 26 at a less demanding signal-to-noise constraint, the fact that the flat distribution of inclinations of trapped interstellar objects would not help with a targeted search, combined with HSC's relatively small field of view, makes LSST the preferred instrument for such a search in the near future.

An `Oumuamua-size object would have an apparent magnitude of $24.5$, LSST's limit, at a distance of $\mathrm{\sim 2 \; AU}$ from the Sun. We will again use the size distribution of Centaurs as an estimate for that of trapped interstellar objects, namely ${n(>D) \propto D^{-1.7}}$, where $D$ is the diameter of the object. The maximum distance at which an object can be observed scales as, ${d \propto D^{\frac{1}{2}}}$. Using the approximations, $d \sim \mathrm{2 \; AU}$ and ${n(>D) \sim 6 \times 10^3}$, for ${D = \mathrm{100 \; m}}$, we express $n(>D)$ as the following function of $d$,

\begin{equation}
n(>D) \sim 6.3 \times 10^4 \cdot (\frac{d}{\mathrm{AU}})^{-3.4} \; \; .
\end{equation}

\begin{figure}
  \centering
  \includegraphics[width=.9\linewidth]{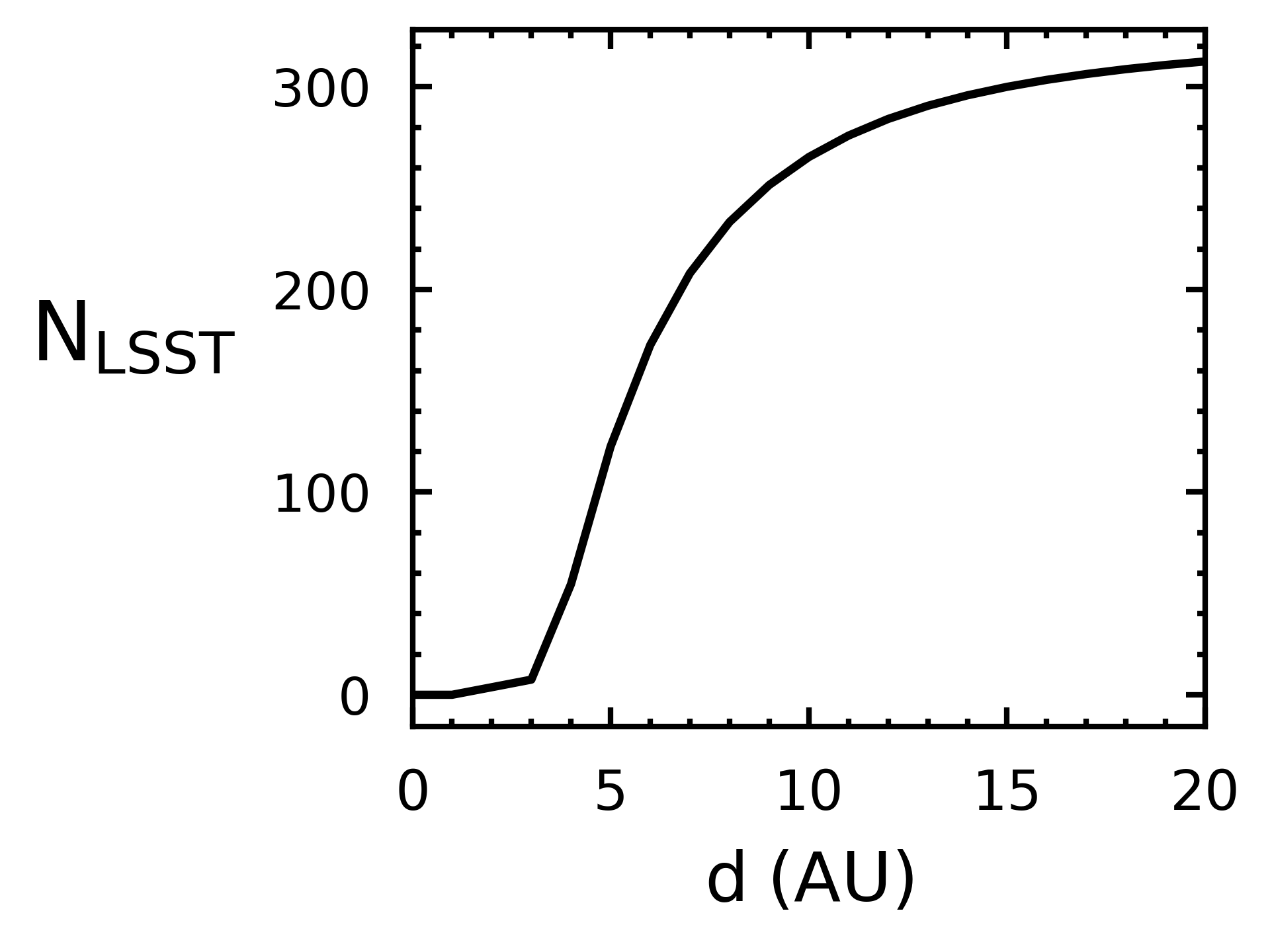}
    \caption{Upper limit of expected LSST detections of trapped interstellar objects in the size range of $\mathrm{100 \; m \; - \; 10 \; km}$, out to a distance, $d$.}
    \label{fig:fig4}
\end{figure}

We then calculate the time-averaged distance of each particle in the population, $d = a(1 + \frac{e^2}{2})$. Finally, we multiply each value of $n(>D)$ by the proportion of the population that is at an average distance $d$, to obtain the number of objects expected to be detectable by LSST as a function of $d$, shown in Fig.~\ref{fig:fig4}. We expect several hundreds of trapped interstellar objects to be detectable by LSST, with diameters ranging from $\sim \mathrm{100 \; m}$ to $\sim \mathrm{10 \; km}$.  The stability lifetime of the trapped orbits is taken into account in this prediction, as it is based on the \citet{Lingam2018} treatment of the steady state population of trapped objects. The lifetime over which an object will receive a ‘kick’ from Jupiter is on the same order as the stability lifetime.

\section{Conclusions}
\label{sec:conclusion}

We simulated the capture of interstellar objects, with initial speeds corresponding to the velocity distribution of stars in the Solar neighborhood, by means of three-body interactions with the Jupiter-Sun system. This was done by first populating the volume surrounding Jupiter with individual particles at their distances of closest approach to the planet. We then integrated the particle trajectories backward in time to determine which ones would be unbound from the Jupiter-Sun system, and subsequently integrated those particles forward in time to find which ones end in bound orbits. We then compared the resulting distributions of orbital elements to those of Centaurs to identify parameters for which interstellar objects should dominate and therefore be easily identifiable. Using these results, we identified known objects that may be of interstellar origin. Finally, we estimated the number of trapped interstellar objects detectable by LSST.

Our calculations indicate that out of the trapped population of $\sim 6 \times 10^3$ `Oumuamua-size interstellar objects, there should be $\sim 4000$ trapped objects in our Solar System identifiable by Centaur-like orbits at $i \gtrsim 48^\circ$. We find that eight known objects have orbital parameters indicating their possible interstellar origin: 2018 WB1, 2013 YG48, 2018 TL6, 2017 SV13, 2013 JD4, 2018 AS18, 2008 WA95, and 2016 WS1, as listed in Table~\ref{tab:1}. Out of the entire population of trapped interstellar objects, we estimate that there are several hundreds of interstellar objects, ranging in diameter from $\sim \mathrm{100 \; m}$ to $\sim \mathrm{10 \; km}$, detectable by LSST.

The orbit of asteroid (514107) 2015 BZ509, which was proposed to have an interstellar origin, is reproducible within the orbital parameter distributions of our trapped population \citep{Namouni2018}. However, its orbital parameters are relatively unlikely, falling in the bottom quartile of both semi-major axis and eccentricity distributions. This could potentially be attributed to detection bias, as objects that are closer in are more likely to be detected by telescopes.

There is strong scientific motivation for investigating interstellar objects, including the potential to gain a deeper understanding of planetary system formation \citep{Seligman2018}. Follow-up observations can help facilitate missions to probe objects of interstellar origin in this orbital parameter space. High-resolution spectroscopy could also be used to measure oxygen isotope ratios for objects with cometary tails in this orbital parameter space, as such ratios are expected to be markedly different for objects of interstellar origin, compared to those which originated within the Solar System \citep{Lingam2018}. Exploration of trapped interstellar objects could potentially help reveal the prospects of life in other star systems as well as extraterrestrial artifacts \citep{Freitas1983, Haqq-Misra2012, Wright2018, 1Lingam2018, Bialy2018}. However, such a search would come with a caveat that most asteroids in the Solar System reside outside the habitable zone.

\section*{Acknowledgements}
We thank Martin Elvis, John Forbes, Manasvi Lingam, Roman Rafikov, and an anonymous referee for helpful comments on the manuscript. We also thank Peter Veres for helping us estimate the population of $\sim \mathrm{100 \; m}$ Centaurs in the Solar System. This research has made use of data and/or services provided by the International Astronomical Union's Minor Planet Center. This work was supported in part by a grant from the Breakthrough Prize Foundation.

\end{document}